\RequirePackage{lineno} 
\documentclass[preprint,superscriptaddress,aps,prl,superscriptaddress,showpacs,preprintnumbers,amsmath,amssymb]{revtex4}
\usepackage{graphicx}
\usepackage{xspace}
\usepackage{epsfig}
\usepackage{epstopdf}
\usepackage{multirow}
\usepackage{dcolumn}  
\usepackage{wasysym}
\usepackage[section]{placeins}
\usepackage{lineno}
\usepackage{subfigure}
\usepackage{color}

\graphicspath{{ps}}

\begin{document}

\preprint{\vbox{ \hbox{   }
							   \hbox{Belle Preprint 2019-06}
							   \hbox{KEK Preprint 2019-4}
}}

\title{ \quad\\[1.0cm]\Large \bf \boldmath Measurement of branching fraction and final-state asymmetry for the $\bar{B}^{0}\to K^{0}_{S}K^{\mp}\pi^{\pm}$ decay}

\noaffiliation


\noaffiliation

\begin{abstract}
We report a measurement of the branching fraction and final-state asymmetry for the $\bar{B}^{0}\to K^{0}_{S}K^{\mp}\pi^{\pm}$ decays. The analysis is based on a data sample of 711 $\rm{fb}^{-1}$ collected at the $\Upsilon(4S)$ resonance with the Belle detector at the KEKB asymmetric-energy $e^{+}e^{-}$ collider. We obtain a branching fraction of $(3.60\pm0.33\pm0.15)\times10^{-6}$ and a final-state asymmetry of $(-8.5\pm8.9\pm0.2)\%$, where the first uncertainties are statistical and the second are systematic. Hints of peaking structures are seen in the differential branching fractions measured as functions of Dalitz variables.

\pacs{14.40.Nd,~13.25.Hw,~13.25.-k,~11.30.Er}
\end{abstract}

\noaffiliation
\affiliation{University of the Basque Country UPV/EHU, 48080 Bilbao}
\affiliation{Beihang University, Beijing 100191}
\affiliation{Brookhaven National Laboratory, Upton, New York 11973}
\affiliation{Budker Institute of Nuclear Physics SB RAS, Novosibirsk 630090}
\affiliation{Faculty of Mathematics and Physics, Charles University, 121 16 Prague}
\affiliation{University of Cincinnati, Cincinnati, Ohio 45221}
\affiliation{Deutsches Elektronen--Synchrotron, 22607 Hamburg}
\affiliation{University of Florida, Gainesville, Florida 32611}
\affiliation{Key Laboratory of Nuclear Physics and Ion-beam Application (MOE) and Institute of Modern Physics, Fudan University, Shanghai 200443}
\affiliation{Justus-Liebig-Universit\"at Gie\ss{}en, 35392 Gie\ss{}en}
\affiliation{Gifu University, Gifu 501-1193}
\affiliation{II. Physikalisches Institut, Georg-August-Universit\"at G\"ottingen, 37073 G\"ottingen}
\affiliation{SOKENDAI (The Graduate University for Advanced Studies), Hayama 240-0193}
\affiliation{Gyeongsang National University, Chinju 660-701}
\affiliation{Hanyang University, Seoul 133-791}
\affiliation{University of Hawaii, Honolulu, Hawaii 96822}
\affiliation{High Energy Accelerator Research Organization (KEK), Tsukuba 305-0801}
\affiliation{J-PARC Branch, KEK Theory Center, High Energy Accelerator Research Organization (KEK), Tsukuba 305-0801}
\affiliation{Forschungszentrum J\"{u}lich, 52425 J\"{u}lich}
\affiliation{IKERBASQUE, Basque Foundation for Science, 48013 Bilbao}
\affiliation{Indian Institute of Technology Bhubaneswar, Satya Nagar 751007}
\affiliation{Indian Institute of Technology Guwahati, Assam 781039}
\affiliation{Indian Institute of Technology Hyderabad, Telangana 502285}
\affiliation{Indian Institute of Technology Madras, Chennai 600036}
\affiliation{Indiana University, Bloomington, Indiana 47408}
\affiliation{Institute of High Energy Physics, Chinese Academy of Sciences, Beijing 100049}
\affiliation{Institute of High Energy Physics, Vienna 1050}
\affiliation{INFN - Sezione di Napoli, 80126 Napoli}
\affiliation{INFN - Sezione di Torino, 10125 Torino}
\affiliation{Advanced Science Research Center, Japan Atomic Energy Agency, Naka 319-1195}
\affiliation{J. Stefan Institute, 1000 Ljubljana}
\affiliation{Institut f\"ur Experimentelle Teilchenphysik, Karlsruher Institut f\"ur Technologie, 76131 Karlsruhe}
\affiliation{Kennesaw State University, Kennesaw, Georgia 30144}
\affiliation{King Abdulaziz City for Science and Technology, Riyadh 11442}
\affiliation{Department of Physics, Faculty of Science, King Abdulaziz University, Jeddah 21589}
\affiliation{Kitasato University, Sagamihara 252-0373}
\affiliation{Korea Institute of Science and Technology Information, Daejeon 305-806}
\affiliation{Korea University, Seoul 136-713}
\affiliation{Kyungpook National University, Daegu 702-701}
\affiliation{LAL, Univ. Paris-Sud, CNRS/IN2P3, Universit\'{e} Paris-Saclay, Orsay}
\affiliation{\'Ecole Polytechnique F\'ed\'erale de Lausanne (EPFL), Lausanne 1015}
\affiliation{P.N. Lebedev Physical Institute of the Russian Academy of Sciences, Moscow 119991}
\affiliation{Faculty of Mathematics and Physics, University of Ljubljana, 1000 Ljubljana}
\affiliation{Ludwig Maximilians University, 80539 Munich}
\affiliation{Luther College, Decorah, Iowa 52101}
\affiliation{Malaviya National Institute of Technology Jaipur, Jaipur 302017}
\affiliation{University of Malaya, 50603 Kuala Lumpur}
\affiliation{University of Maribor, 2000 Maribor}
\affiliation{Max-Planck-Institut f\"ur Physik, 80805 M\"unchen}
\affiliation{School of Physics, University of Melbourne, Victoria 3010}
\affiliation{University of Mississippi, University, Mississippi 38677}
\affiliation{University of Miyazaki, Miyazaki 889-2192}
\affiliation{Moscow Physical Engineering Institute, Moscow 115409}
\affiliation{Moscow Institute of Physics and Technology, Moscow Region 141700}
\affiliation{Graduate School of Science, Nagoya University, Nagoya 464-8602}
\affiliation{Kobayashi-Maskawa Institute, Nagoya University, Nagoya 464-8602}
\affiliation{Universit\`{a} di Napoli Federico II, 80055 Napoli}
\affiliation{Nara Women's University, Nara 630-8506}
\affiliation{National Central University, Chung-li 32054}
\affiliation{National United University, Miao Li 36003}
\affiliation{Department of Physics, National Taiwan University, Taipei 10617}
\affiliation{H. Niewodniczanski Institute of Nuclear Physics, Krakow 31-342}
\affiliation{Niigata University, Niigata 950-2181}
\affiliation{Novosibirsk State University, Novosibirsk 630090}
\affiliation{Osaka City University, Osaka 558-8585}
\affiliation{Pacific Northwest National Laboratory, Richland, Washington 99352}
\affiliation{Panjab University, Chandigarh 160014}
\affiliation{Peking University, Beijing 100871}
\affiliation{University of Pittsburgh, Pittsburgh, Pennsylvania 15260}
\affiliation{Punjab Agricultural University, Ludhiana 141004}
\affiliation{Theoretical Research Division, Nishina Center, RIKEN, Saitama 351-0198}
\affiliation{University of Science and Technology of China, Hefei 230026}
\affiliation{Seoul National University, Seoul 151-742}
\affiliation{Showa Pharmaceutical University, Tokyo 194-8543}
\affiliation{Soongsil University, Seoul 156-743}
\affiliation{University of South Carolina, Columbia, South Carolina 29208}
\affiliation{Stefan Meyer Institute for Subatomic Physics, Vienna 1090}
\affiliation{Sungkyunkwan University, Suwon 440-746}
\affiliation{School of Physics, University of Sydney, New South Wales 2006}
\affiliation{Department of Physics, Faculty of Science, University of Tabuk, Tabuk 71451}
\affiliation{Tata Institute of Fundamental Research, Mumbai 400005}
\affiliation{Department of Physics, Technische Universit\"at M\"unchen, 85748 Garching}
\affiliation{Toho University, Funabashi 274-8510}
\affiliation{Department of Physics, Tohoku University, Sendai 980-8578}
\affiliation{Earthquake Research Institute, University of Tokyo, Tokyo 113-0032}
\affiliation{Department of Physics, University of Tokyo, Tokyo 113-0033}
\affiliation{Tokyo Institute of Technology, Tokyo 152-8550}
\affiliation{Tokyo Metropolitan University, Tokyo 192-0397}
\affiliation{Virginia Polytechnic Institute and State University, Blacksburg, Virginia 24061}
\affiliation{Wayne State University, Detroit, Michigan 48202}
\affiliation{Yamagata University, Yamagata 990-8560}
\affiliation{Yonsei University, Seoul 120-749}
  \author{Y.-T.~Lai}\affiliation{High Energy Accelerator Research Organization (KEK), Tsukuba 305-0801} 
  \author{I.~Adachi}\affiliation{High Energy Accelerator Research Organization (KEK), Tsukuba 305-0801}\affiliation{SOKENDAI (The Graduate University for Advanced Studies), Hayama 240-0193} 
  \author{H.~Aihara}\affiliation{Department of Physics, University of Tokyo, Tokyo 113-0033} 
  \author{S.~Al~Said}\affiliation{Department of Physics, Faculty of Science, University of Tabuk, Tabuk 71451}\affiliation{Department of Physics, Faculty of Science, King Abdulaziz University, Jeddah 21589} 
  \author{D.~M.~Asner}\affiliation{Brookhaven National Laboratory, Upton, New York 11973} 
  \author{H.~Atmacan}\affiliation{University of South Carolina, Columbia, South Carolina 29208} 
  \author{V.~Aulchenko}\affiliation{Budker Institute of Nuclear Physics SB RAS, Novosibirsk 630090}\affiliation{Novosibirsk State University, Novosibirsk 630090} 
  \author{T.~Aushev}\affiliation{Moscow Institute of Physics and Technology, Moscow Region 141700} 
  \author{V.~Babu}\affiliation{Tata Institute of Fundamental Research, Mumbai 400005} 
  \author{I.~Badhrees}\affiliation{Department of Physics, Faculty of Science, University of Tabuk, Tabuk 71451}\affiliation{King Abdulaziz City for Science and Technology, Riyadh 11442} 
% ...
\collaboration{The Belle Collaboration}

\maketitle

Three-body charmless hadronic $B$ decays are sensitive to $CP$ violation localized in their Dalitz plane~\cite{cp_dalitz,cp_dalitz_1}. Charmless $B$ decays are suppressed in the standard model (SM), and decays with an even number of kaons, such as $\bar{B}^{0}\to K^{0}_{S}K^{\mp}\pi^{\pm}$~\cite{charge_conjugate}, have a smaller decay rate compared to those with an odd number of kaons.
These proceed via $b\to u$ trees and $W$-exchange, and via a $b\to d$ penguin process with a virtual loop; the latter provides an opportunity to search for physics beyond the SM since new heavy particles may cause deviations from SM predictions.

Previous measurements by the BABAR~\cite{babar_Btokpik0,babar_kstk0} and LHCb~\cite{LHCb_Btokpik0_latest,LHCb_kstk,LHCb_kstks} experiments found hints of structures in the low $K^{-}\pi^{+}$ and  $K^{-}K^{0}_{S}$ mass regions that have highly asymmetric helicity angular distributions.
However, the yields are not sufficient to draw firm conclusions with a full Dalitz analysis. Similar studies on $B^{+}\to K^{+}K^{-}\pi^{+}$ were performed by Belle~\cite{kkpi_1}, BABAR~\cite{kkpi_2}, and LHCb~\cite{kkpi_3,kkpi_4}, in which strong evidence of localized $CP$ violation was found in the low $M_{K^{+}K^{-}}$ region.


By using the full data set of Belle, we expect to measure the branching fraction and final-state asymmetry of $\bar{B}^{0}\to K^{0}_{S}K^{\mp}\pi^{\pm}$ decays more precisely. Using the charges of final-state particles, the latter is defined as 
\begin{equation}
\mathcal{A}=\frac{N(K^{0}_{S}K^{-}\pi^{+})-N(K^{0}_{S}K^{+}\pi^{-})}{N(K^{0}_{S}K^{-}\pi^{+})+N(K^{0}_{S}K^{+}\pi^{-})},
\end{equation} 
where $N$ denotes the measured signal yield of the corresponding $B$ final states, and $N(K^{0}_{S}K^{-}\pi^{+})=N(B^{0}\to K^{0}_{S}K^{-}\pi^{+})+N(\bar{B}^{0}\to K^{0}_{S}K^{-}\pi^{+})$. Here $\mathcal{A}$ is distinct from the direct $CP$ asymmetry ($\mathcal{A}_{CP}$); rather it
is an asymmetry between the decay final states of $K^{0}K^{-}\pi^{+}$ and $\bar{K}^{0}K^{+}\pi^{-}$ where $K^{0}(\bar{K}^{0})$ leads to a $K^{0}_{S}$. 
We measure this quantity since it can be more precisely determined than $\mathcal{A}_{CP}$ for this decay mode. This is the first measurement of such an asymmetry for the three-body $\bar{B}^{0}\to K^{0}_{S}K^{\mp}\pi^{\pm}$ decay.
In addition, we use the $_s\mathcal{P}lot$~\cite{splot} method to obtain background-subtracted yields for the Dalitz variables $M_{K^{-}\pi^{+}}$, $M_{\pi^{+}K^{0}_{S}}$, and $M_{K^{-}K^{0}_{S}}$, and hence determine their differential branching fractions. The total branching fraction is extracted by integrating the differential branching fraction.

Our measurement is based on a data sample of 711 $\rm{fb^{-1}}$, corresponding to $772\times10^{6}$ $B\bar{B}$ pairs, collected with the Belle detector~\cite{Belle} operating at the KEKB asymmetric-energy $e^{+} e^{-}$ collider~\cite{KEKB}. The Belle detector is a large-solid-angle magnetic spectrometer that consists of a silicon vertex detector (SVD), a 50-layer central drift chamber (CDC), an array of aerogel threshold Cherenkov counters (ACC), a barrel-like arrangement of time-of-flight scintillation counters (TOF) and an electromagnetic calorimeter comprised of CsI(Tl) crystals, all located inside a superconducting solenoid that provides a 1.5~T magnetic field. An iron flux-return yoke located outside the solenoid is instrumented to detect $K_L^0$ mesons and muons. 
The detector is described in detail elsewhere~\cite{Belle}. 

This analysis uses two data sets with different inner-detector configurations.
The first data set of 140 $\rm{fb}^{-1}$ was collected with a beam pipe of radius 2.0 cm and with 3 layers of SVD, while the second data set of 571 $\rm{fb}^{-1}$ was recorded with a beam pipe of radius 1.5~cm and 4 layers of SVD \cite{svd2}. 
Large samples of Monte Carlo (MC) events for signal and backgrounds are generated with EvtGen~\cite{ref:EvtGen} and subsequently simulated with GEANT3~\cite{geant} with the configurations of the Belle detector. These samples are used to obtain the expected distributions of various physical quantities for signal and backgrounds, to optimize the selection criteria as well as to determine the signal detection efficiency.

The selection criteria for the final-state charged particles in the $\bar{B}^{0}\to K^{0}_{S}K^{\mp}\pi^{\pm}$ reconstruction are based on information obtained from the tracking systems (SVD and CDC) and the charged-hadron identification (PID) systems, namely the CDC, ACC, and TOF. The charged kaons and pions are required to have an impact parameter within $\pm 0.2$ cm of the interaction point (IP) in the transverse plane, and within $\pm 5.0$ cm along the $e^{+}$ beam direction.
The likelihood values of each track for kaon and pion hypotheses ($L_{K}$ and $L_{\pi}$) are determined from the information provided by the PID system. A track is identified as a kaon if $L_{K}/(L_{K}+L_{\pi}) > 0.6$ otherwise it is treated as a pion. The efficiency for identifying
a pion (kaon) is about 88\% (86\%), which depends on the momenta of the track, while the probability for a pion or a kaon to be misidentified is less than 10\%. The efficiency and misidentification probabilities are averaged over the momentum of the final-state particles. The $K^{0}_{S}$ candidates are reconstructed via the $K^{0}_{S}\to\pi^{+}\pi^{-}$ decay, and the identification is enhanced by selecting on the output of a neural network (NN)~\cite{NN}, which combines seven kinematic variables of the $K^{0}_{S}$~\cite{nisKs}. The invariant mass of the $K^{0}_{S}$ candidates is required to be within $\pm10$ MeV/$c^{2}$ of the world average, which corresponds to about three times the resolution. The $K^{0}_{S}\to \pi^{+}\pi^{-}$ vertex fit is required to converge with a goodness-of-fit value $(\chi^{2})$ less than 20.

$B$ mesons are identified with two kinematic variables calculated in the center-of-mass (CM) frame:  the beam-energy-constrained-mass $M_{\mathrm{bc}}\equiv \sqrt{E^{2}_{\mathrm{beam}}/c^{4}-|\vec{p}_{B}/c|^{2}}$, and the energy difference $\Delta E\equiv E_{B}-E_{\mathrm{beam}}$, where $E_{\mathrm{beam}}$ is the beam energy, 
and $\vec{p}_B$ ($E_B$) is the momentum (energy) of the reconstructed $B$ meson. The $B$ candidates are required to have $M_{\rm bc} >$ 5.255 GeV/$c^{2}$ and $ |\Delta E| <$ 0.15 GeV, and the signal region is defined as 5.272 GeV/$c^{2}$ $< M_{\rm bc} <$ 5.288 GeV/$c^{2}$ and $ |\Delta E| <$ 0.05 GeV. We require a successful vertex fit for $\bar{B}^{0}\to K^{0}_{S}K^{\mp}\pi^{\pm}$ candidates, where the $K^{0}_{S}$ trajectory is included in the fit, with $\chi^{2} <$ 100. We find that 9\% of events have more than one $B$ candidate. In such cases, we choose the candidate with the smallest $\chi^{2}$ value. According to simulation, our best entry selection method chooses the correct candidate in 99\% of cases.

The dominant background arises from the continuum $e^{+}e^{-} \to q\bar{q}~(q = u,d,s,c)$ process. To suppress this, we construct a Fisher discriminant~\cite{Fisher} from 17 modified Fox-Wolfram moments ~\cite{KSFW}. To further improve the distinguishing power, we combine the output of the Fisher discriminant with four more variables in a NN. These are: the cosine of the angle between the reconstructed $B$ flight direction and the beam direction in the CM frame, the offset along the $z$ axis between the vertex of the reconstructed $B$ and the vertex formed by the remaining tracks, the cosine of the angle between the thrust axis~\cite{thrust} of the reconstructed $B$ and that of the rest of the event in the CM frame, and a $B$ meson flavor tagging quality variable. The NN is trained with signal and continuum MC samples. The NN output ($C_{\mathrm{NN}}$) ranges from $-1$ to 1, and it is required to be greater than 0.7. This removes 93\% of the continuum background while 82\% of the signal is retained. We transform $C_{\mathrm{NN}}$ to $C^{'}_{\mathrm{NN}}\equiv \mathrm{log}(\frac{C_{\mathrm{NN}}-C^{\mathrm{min}}_{\mathrm{NN}}}{C^{\mathrm{max}}_{\mathrm{NN}}-C_{\mathrm{NN}}})$, where $C^{\mathrm{min}}_{\mathrm{NN}}$ is 0.7 and $C^{\mathrm{max}}_{\mathrm{NN}}$ is the maximum value of $C_{\mathrm{NN}}$.

Background events from $B$ decays mediated via the $b \to c$ transition (generic $B$ decays) may have peaking structures in the signal region. They are mainly due to the decays with two-body final states of $D$ mesons and $J/\psi$, e.g., $D^{0}\to K^{-}\pi^{+}$,  $D^{-}\to K^{-}K^{0}_{S}$, $D^{-}_{s}\to K^{-}K^{0}_{S}$, $J/\psi\to e^{+}e^{-}$, and $J/\psi\to \mu^{+}\mu^{-}$. These decays can be identified by peaks at the nominal $D$ and $J/\psi$ masses in the distributions of the invariant masses of two of the final-state particles ($M_{K^{-}\pi^{+}}$, $M_{\pi^{+}K^{0}_{S}}$, $M_{K^{-}K^{0}_{S}}$, where we allow for a change in the mass hypothesis of a charged kaon or pion). We exclude events within $\pm4\sigma$ of the nominal mass of the peaking structures to suppress the contributions from $D$ mesons and $J/\psi$.

The rare $B$ background coming from $b\to u, d, s$ transitions is studied with a large MC sample in which the branching fractions are much larger than the measured or expected value. Two modes are found to have peaks near the $\Delta E$ signal region: $B^{0}\to K^{-}K^{+}K^{0}_{S}$ and $B^{0}\to \pi^{-}\pi^{+}K^{0}_{S}$, including their intermediate resonant modes. The remaining rare $B$ events have a relatively flat $\Delta E$ distribution.

The signal yield and $\mathcal{A}$ are extracted 
from a three-dimensional extended unbinned maximum likelihood fit, with the likelihood defined as 
\begin{equation}
\mathcal{L}=\frac{e^{-\sum_{j}N_{j}}}{N!}\prod^{N}_{i=1}\left(\sum_{j}N_{j}P^{i}_{j}\right),
\end{equation}
where,
\begin{equation}
P^{i}_{j}=\frac{1}{2}(1-q^{i}\cdot\mathcal{A}_{j})\times P_{j}(M_{\rm bc}^{i},\Delta E^{i},C^{'i}_{\mathrm{NN}}),
\end{equation}
$N$ is the total number of candidate events, $N_{j}$ is the number of  
events in category $j$, $i$ denotes the event index, $q^{i}$ is the charge of the $K^{\pm}$ in the $i$-th event, $\mathcal{A}_{j}$ is the value of final-state asymmetry of the $j$-th category, $P_{j}$ represents the value of the corresponding three-dimensional probability density function (PDF), and $M_{\rm bc}^{i}$, $\Delta E^{i}$, and $C^{'i}_{\mathrm{NN}}$ are the $M_{\rm bc}$, $\Delta E$, and $C^{'}_{\mathrm{NN}}$ values of the $i$-th event, respectively.

With all the selection criteria applied, the signal MC sample contains 98\% of the correctly-reconstructed signal $B$ events (`true' signal) and 2\% self-crossfeed (scf) events.
In the fit, the ratio of scf to true signal events is fixed. The signal yield ($N_{\mathrm{sig}}$) is the combined yield of the true signal PDF and the scf PDF. 
In addition to the signal, five more categories are included in the fit: continuum background, generic $B\bar{B}$ background, $B^{0}\to K^{-}K^{+}K^{0}_{S}$, $B^{0}\to \pi^{-}\pi^{+}K^{0}_{S}$, and the remaining rare $B$ background. The true signal PDF is described by the product of a sum of two Gaussian functions in $M_{\rm bc}$, a sum of three Gaussian functions in $\Delta E$, and an asymmetric Gaussian function in $C^{'}_{\mathrm{NN}}$. These signal PDF shapes are calibrated including possible data-MC differences obtained from a study of a control mode: $B^{0}\to D^{-}\pi^{+}$ with $D^{-}\to K^{0}_{S}\pi^{-}$. The continuum background PDF is described by the product of an ARGUS function~\cite{argus} in $M_{\mathrm{bc}}$, a second-order polynomial in $\Delta E$, and a combination of a Gaussian and an asymmetric Gaussian function in $C^{'}_{\mathrm{NN}}$. The shape parameters of the continuum background PDF are free in the data fit, except for the ARGUS endpoint which is fixed to 5.2892 GeV/$c^{2}$. For the other contributions (scf, generic $B$, $B^{0}\to K^{-}K^{+}K^{0}_{S}$, $B^{0}\to \pi^{-}\pi^{+}K^{0}_{S}$, and rare $B$), their PDFs are described by a smoothed histogram in $\Delta E$ and $M_{\mathrm{bc}}$, and an asymmetric Gaussian function in $C^{'}_{\mathrm{NN}}$ whose shape is based on MC. The yield of each category is floated. Except for the signal, $\mathcal{A}$ is fixed to zero for the other background categories.

\begin{figure}[htb]
\centering
\subfigure[$\Delta E$]{
\includegraphics[width=0.3\textwidth]{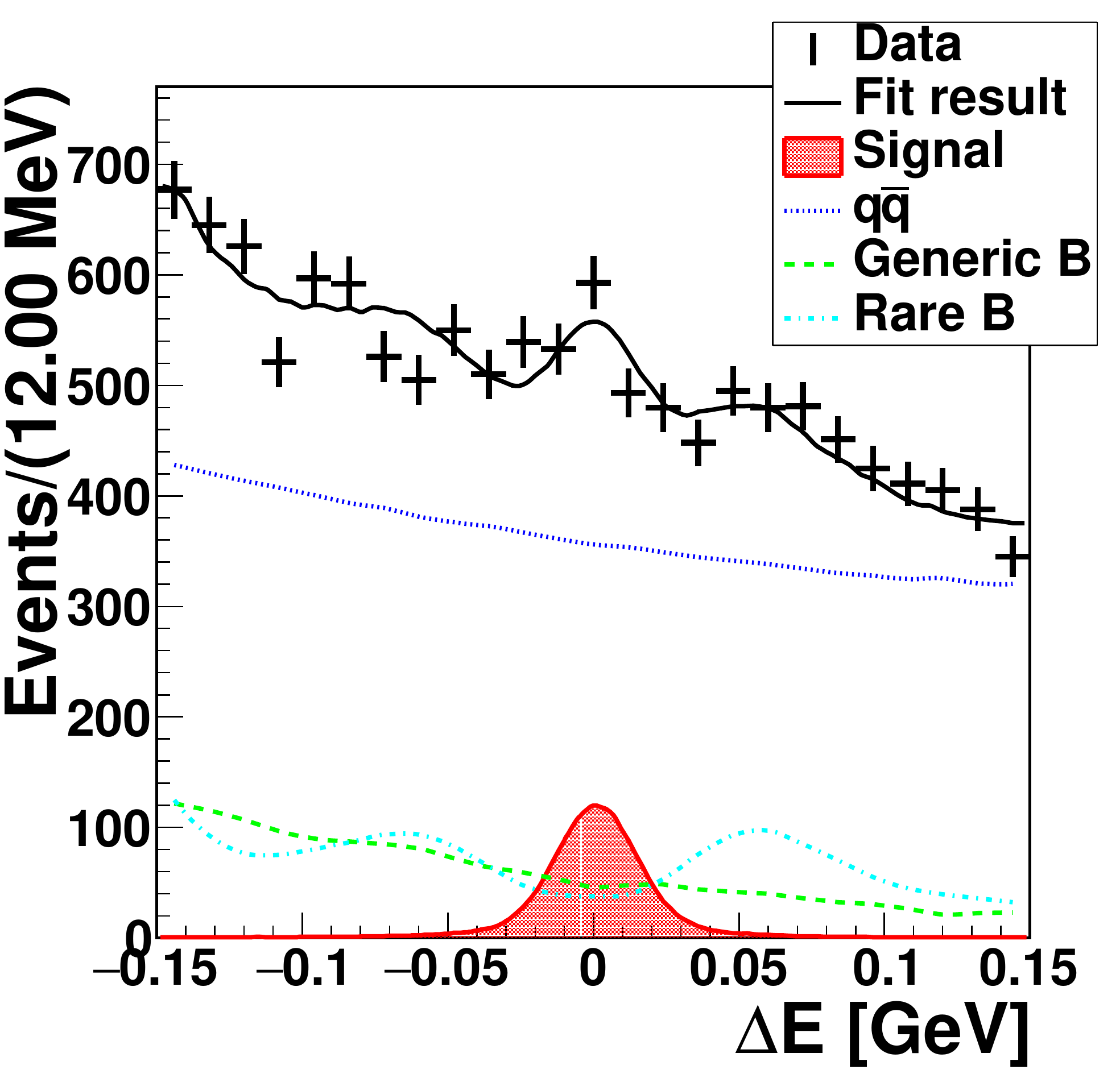}}
\subfigure[$M_{\mathrm{bc}}$]{
\includegraphics[width=0.3\textwidth]{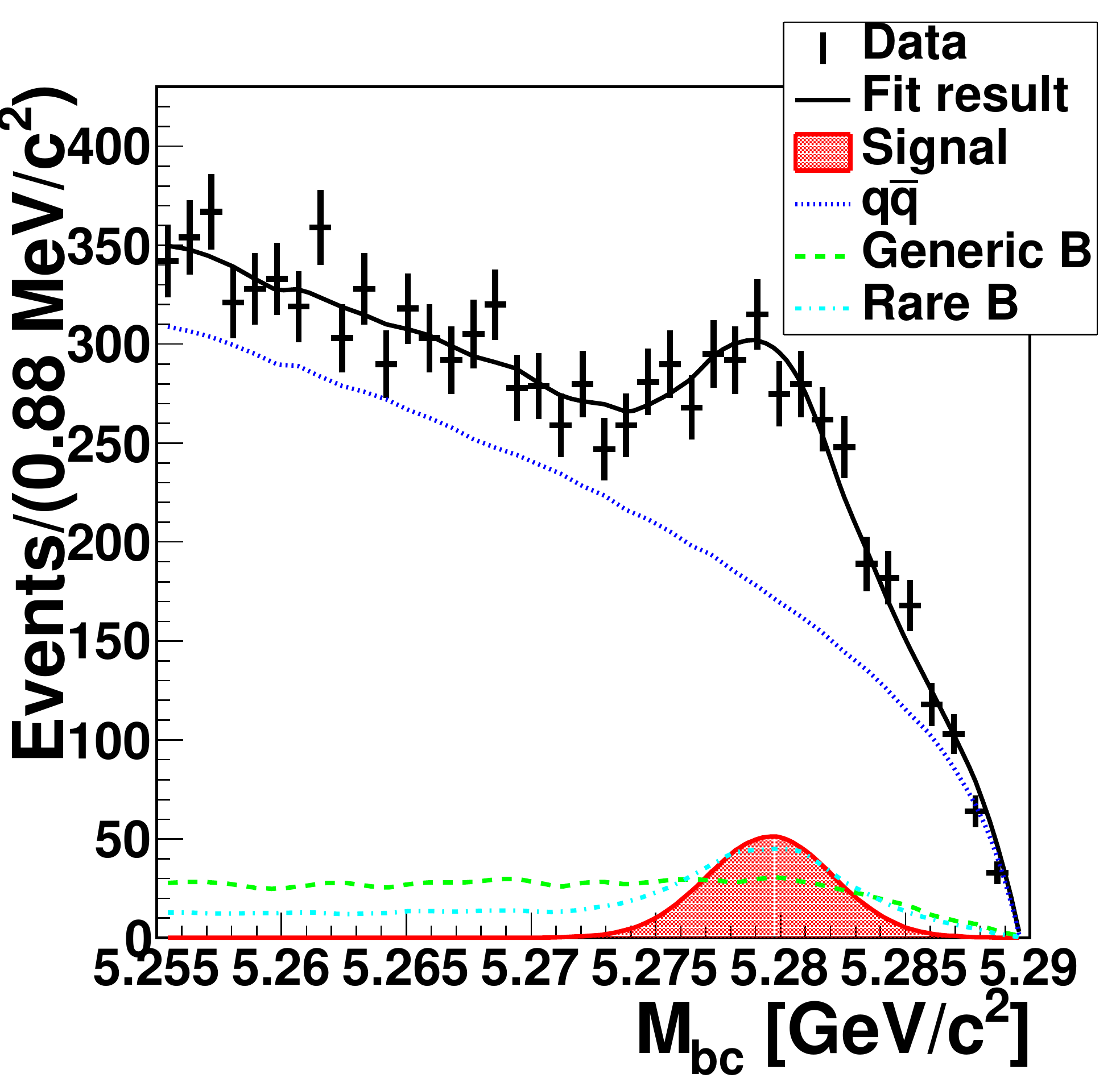}}
\subfigure[$C^{'}_{NN}$]{
\includegraphics[width=0.3\textwidth]{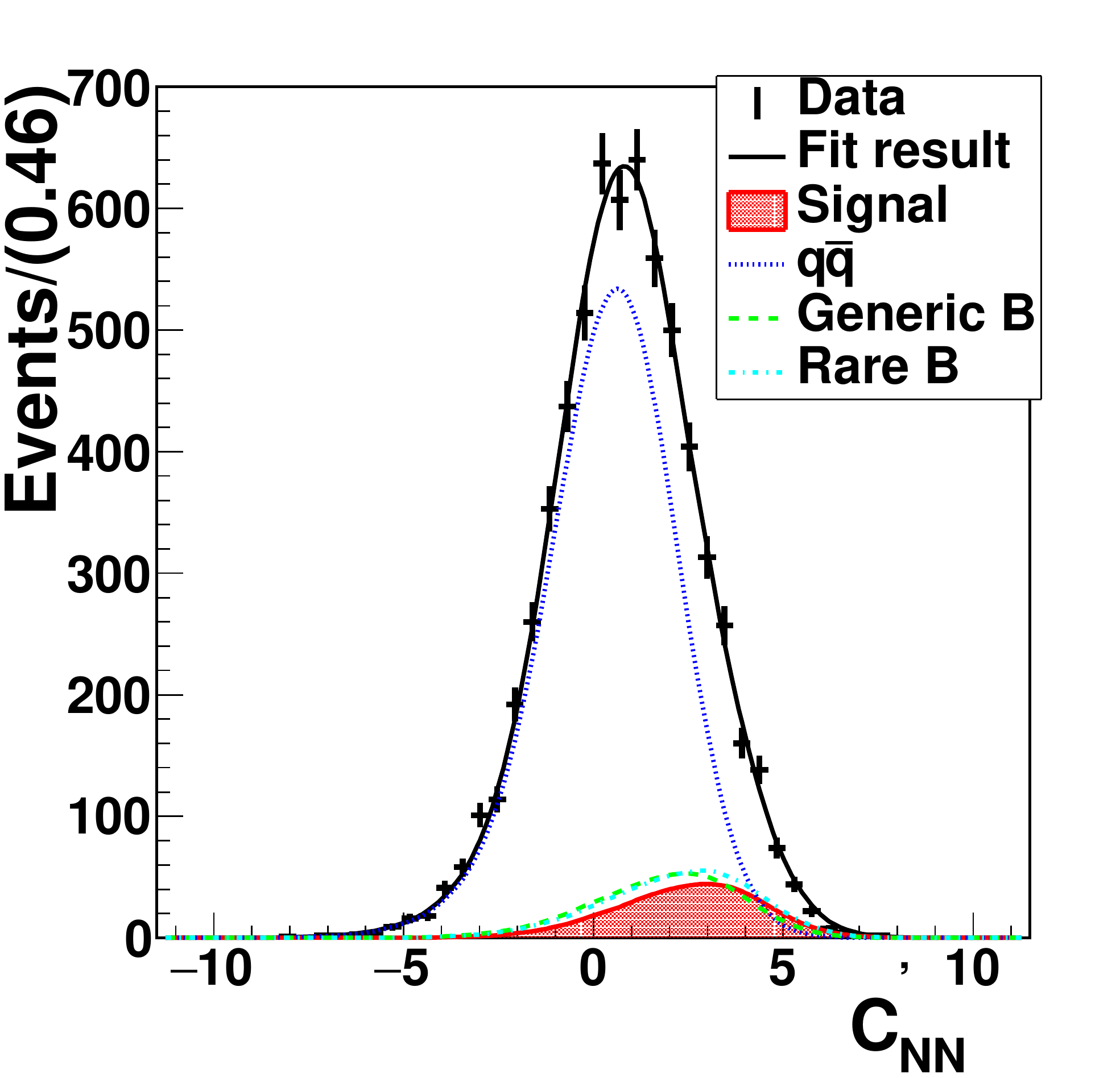}}
\caption{Signal-enhanced projections of the fit results of $\bar{B}^{0}\to K^{0}_{S}K^{\mp}\pi^{\pm}$ decay on $\Delta E$, $M_{\mathrm{bc}}$, and $C^{'}_{\mathrm{NN}}$. (a) $\Delta E$ in  5.272 GeV/$c^{2}<M_{\mathrm{bc}}<$5.288 GeV/$c^{2}$ and $0<C^{'}_{\mathrm{NN}}<5$. (b) $M_{\mathrm{bc}}$ in $|\Delta E|<$ 0.05 GeV and $0<C^{'}_{\mathrm{NN}}<5$. (c) $C^{'}_{\mathrm{NN}}$ in $|\Delta E|<$ 0.05 GeV and 5.272 GeV/$c^{2}<M_{\mathrm{bc}}<$5.288 GeV/$c^{2}$.}
\label{fg:fit} 
\end{figure}

The signal-enhanced projections of the fit are shown in Fig.~\ref{fg:fit}. We obtain a signal yield of $490^{+46}_{-45}$ with a statistical significance of 13 standard deviations, and an $\mathcal{A}$ of $(-8.5\pm8.9)\%$. The significance is defined as $\sqrt{-2\rm{ln}(\mathcal{L}_{0}/\mathcal{L}_{max})}$, where $\mathcal{L}_{0}$ and $\mathcal{L}_{\rm{max}}$ are the likelihood values obtained by the fit with and without the signal yield fixed to zero, respectively. 

The branching fraction is calculated using 
\begin{equation}
\mathcal{B}=\frac{N_{\rm{sig}}}{\epsilon \times \eta \times N_{B\bar{B}}},
\end{equation}
where $N_{\rm{sig}}$, $N_{B\bar{B}}$, $\epsilon$, and $\eta$ are the fitted signal yield, 
the number of $B\bar{B}$ pairs ($=772\times10^{6}$), the reconstruction efficiency of the signal, and the efficiency calibration factor, respectively. We assume that charged and neutral $B\bar{B}$ pairs are produced equally at the $\Upsilon(4S)$. The reconstruction efficiency for the signal ($\epsilon$) is $(26.7\pm0.03)\%$ which is determined by MC only and with all the selection criteria applied. The last quantity contains calibrations due to various systematic effects $\eta=\eta_{K}\times\eta_{\pi}\times\eta_{\textrm{NN}}\times\eta_{\textrm{fit}}$, where $\eta_{K}(=0.9948\pm0.0083$) and  $\eta_{\pi}(=0.9512\pm0.0079$) are the corrections due to $K^{\pm}$ and $\pi^{\pm}$ identification with requirements on $L_{K}$ and $L_{\pi}$, and are obtained by a control sample study of $D^{*+}\to D^{0}\pi^{+}$ with $D^{0}\to K^{-}\pi^{+}$, $\eta_{\textrm{NN}}(=0.9897\pm0.0208)$ is due to the requirement on $C_{\mathrm{NN}}$ and is obtained from $B^{0}\to D^{-}\pi^{+}$ data with a $D^{-}\to K^{0}_{S}\pi^{-}$ control sample study, and $\eta_{\textrm{fit}}(=1.022\pm0.004)$ is due to fit bias and is obtained from an ensemble test on the fitter.

\begin{figure}[htb]
\centering
\includegraphics[width=0.4\textwidth]{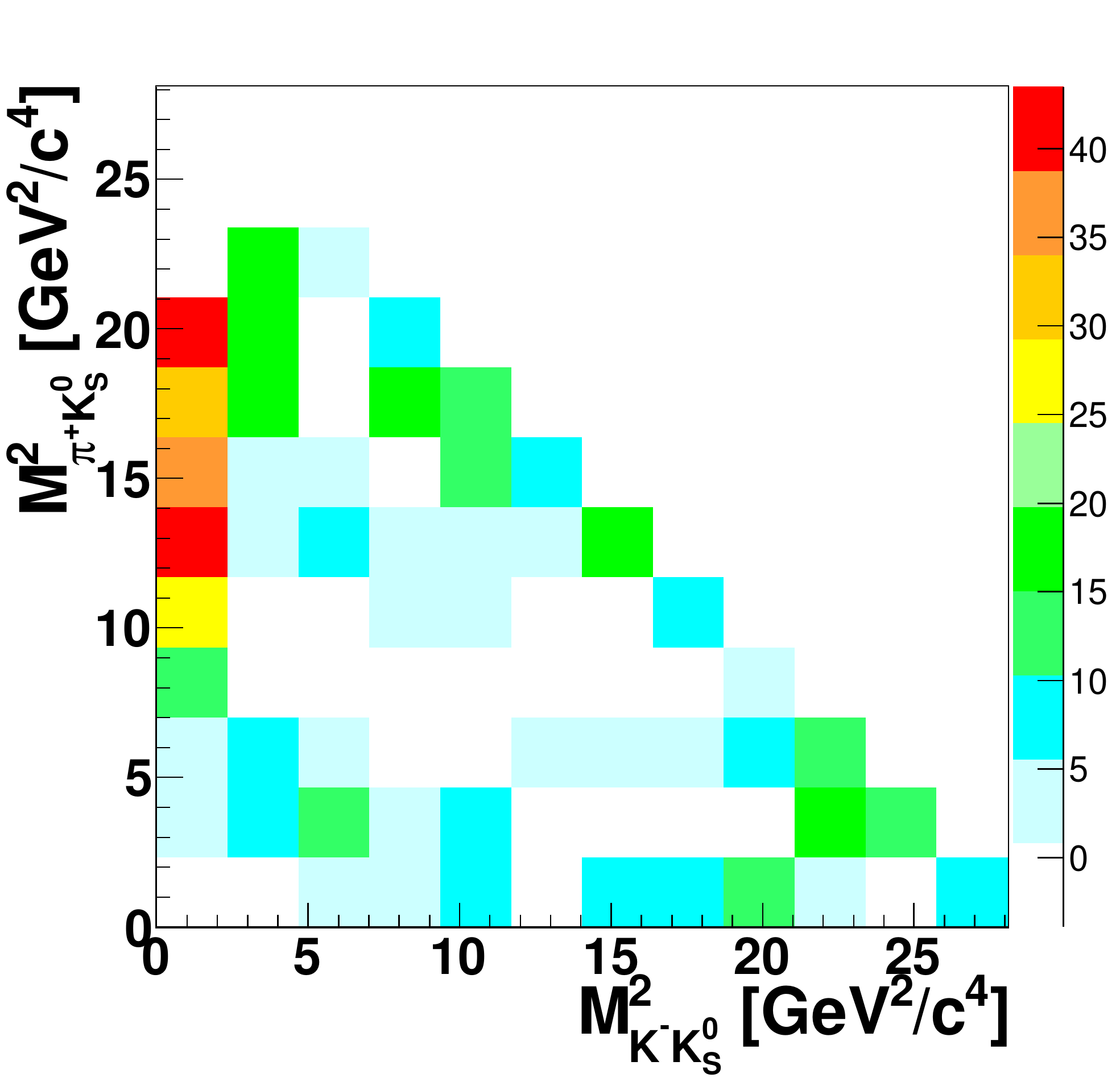}
\caption{Background-subtracted Dalitz plot of the $\bar{B}^{0}\to K^{0}_{S}K^{\mp}\pi^{\pm}$ decay.}
\label{fg:dalitz} 
\end{figure}

\begin{figure}[htb]
\centering
\subfigure[$M_{K^{-}\pi^{+}}$]{
\includegraphics[width=0.3\textwidth]{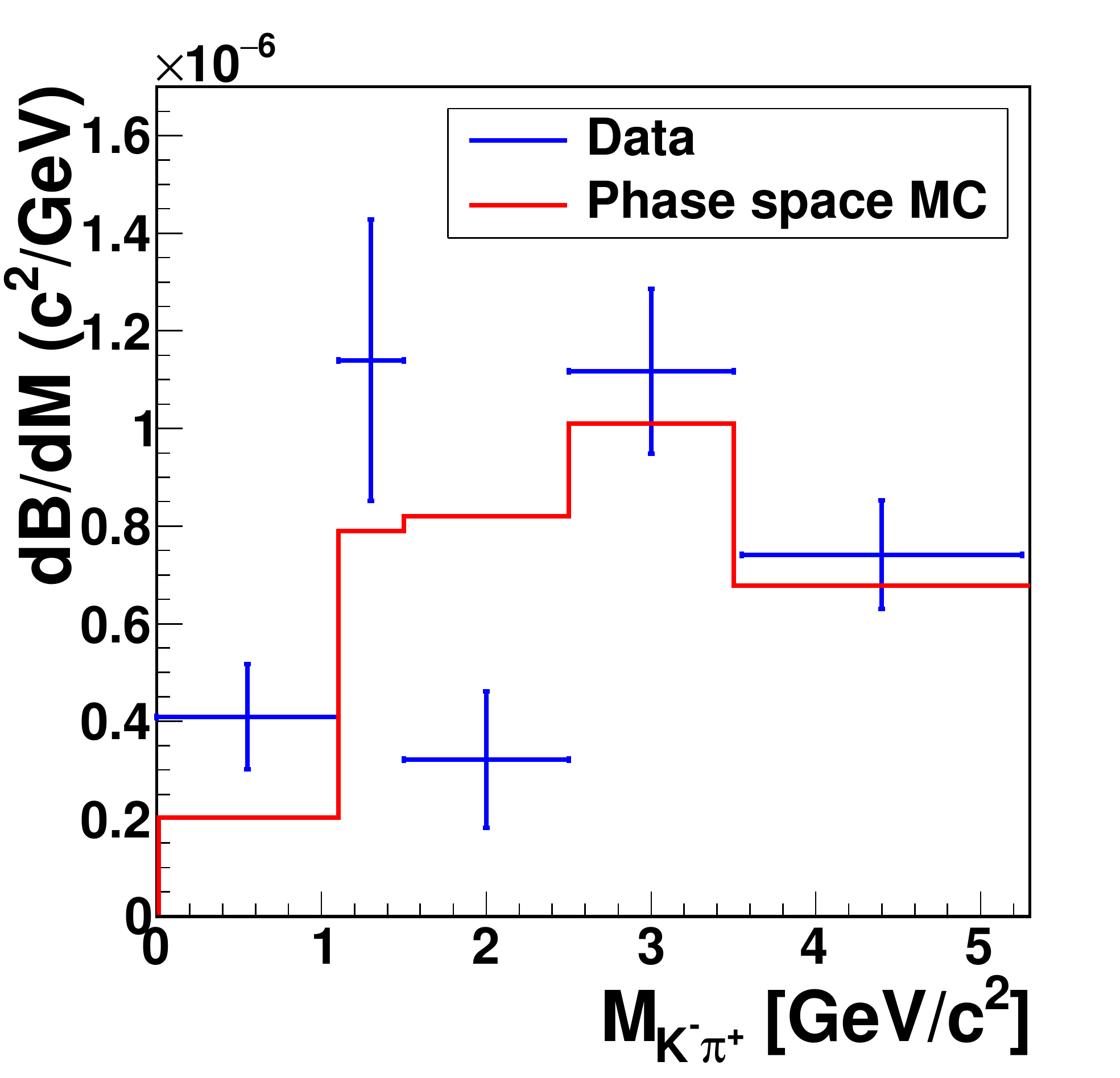}}
\subfigure[$M_{K^{-}K^{0}_{S}}$]{
\includegraphics[width=0.3\textwidth]{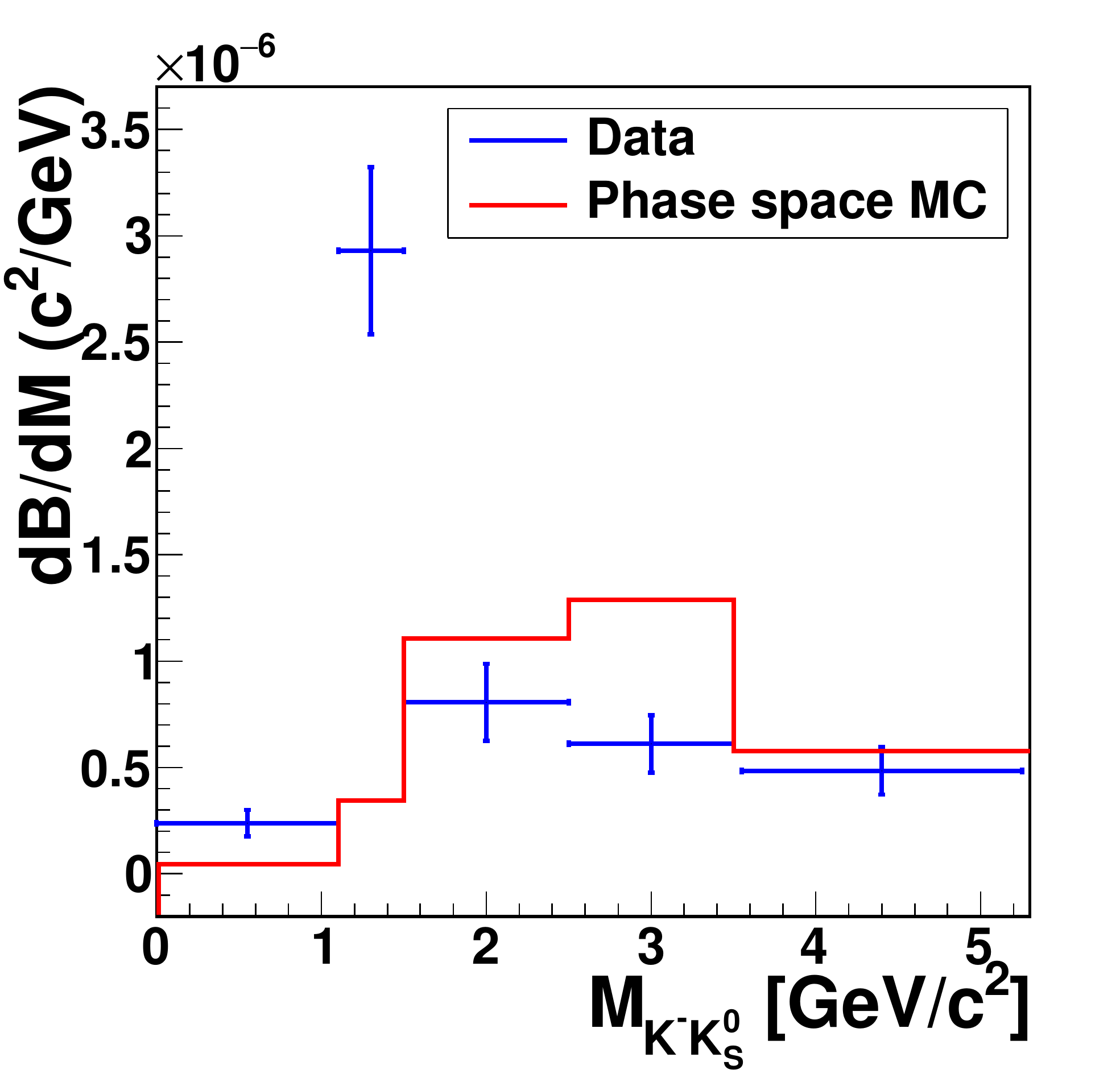}}
\subfigure[$M_{\pi^{+}K^{0}_{S}}$]{
\includegraphics[width=0.3\textwidth]{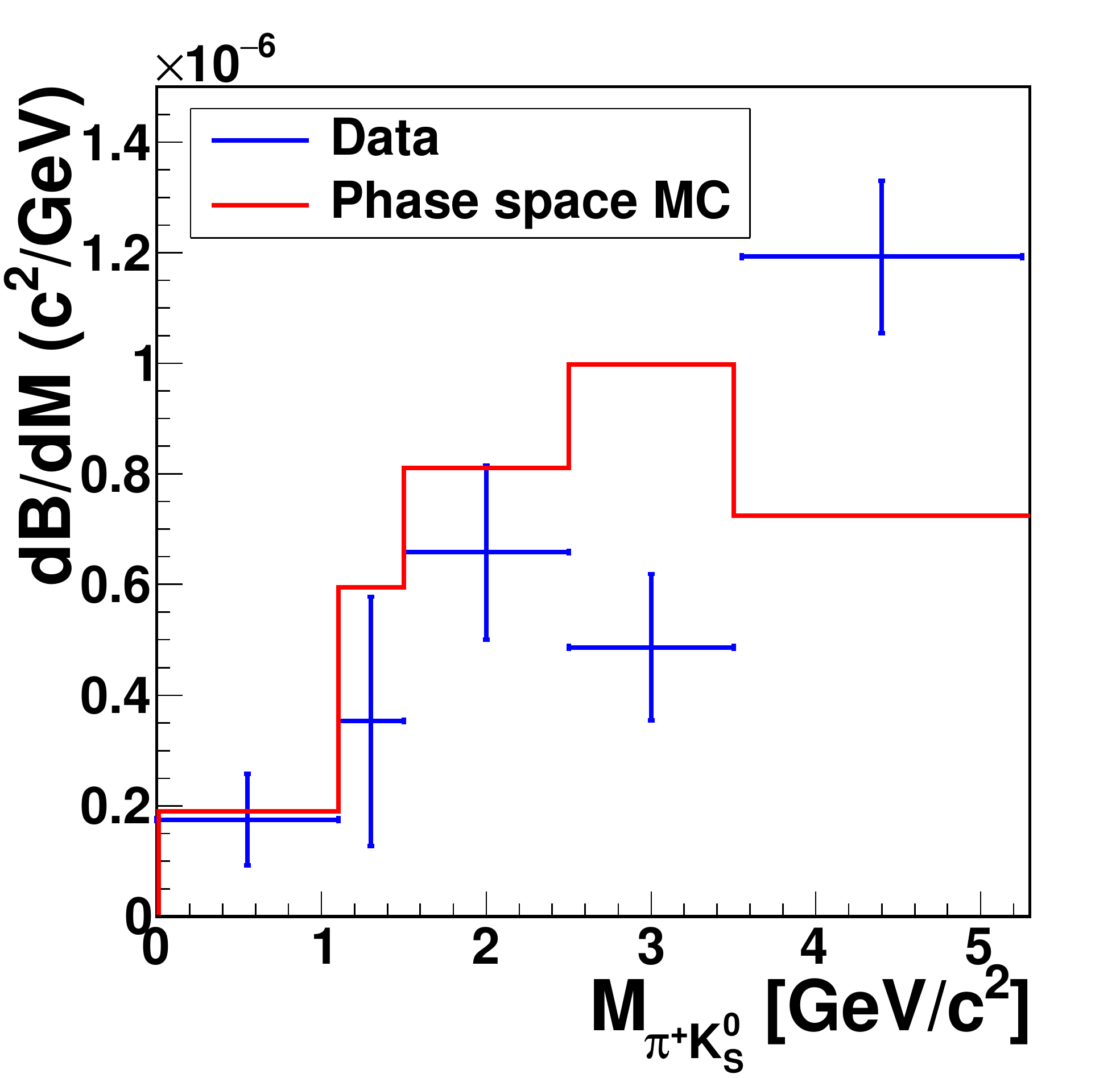}}
\caption{Differential branching fraction as functions of $M_{K^{-}\pi^{+}}$, $M_{K^{-}K^{0}_{S}}$, and $M_{\pi^{+}K^{0}_{S}}$. The points with blue error bars is the data result. The red histogram is obtained by using a signal MC sample with a 3-body phase space decay model.}
\label{fg:dbf} 
\end{figure}

Figure~\ref{fg:dalitz} shows the background-subtracted Dalitz plot obtained with the $_s\mathcal{P}lot$ method. Structures around the regions $M^{2}_{K^{-}K^{0}_{S}}<$ 2 GeV$^{2}$/$c^{4}$ and 7 GeV$^{2}$/$c^{4}$ $<M^{2}_{\pi^{+}K^{0}_{S}} <$ 23 GeV$^{2}$/$c^{4}$ are visible. We also obtain background-subtracted distributions after separating into five bins, and then calculate the differential branching fractions as functions of the three Dalitz variables with the yield and reconstruction efficiency within each bin. We use a similar binning scheme as the one in the $B^{+}\to K^{+} K^{-} \pi^{+}$ measurement at Belle~\cite{kkpi_1}. Figure~\ref{fg:dbf} shows the differential branching fractions as functions of the three Dalitz variables including comparison to the MC with a three-body phase space decay model. Large deviations from phase space expectations are found in the second bin (around 1.2 GeV/$c^{2}$) of the $M_{K^{-}K^{0}_{S}}$ spectrum and at the fourth and fifth bin (around 3.0 GeV/$c^{2}$ - 4.2 GeV/$c^{2}$) in the $M_{\pi^{+}K^{0}_{S}}$ spectrum. In addition, no obvious structure is observed in the low-mass regions of both $M_{K^{-}\pi^{+}}$ and $M_{\pi^{+}K^{0}_{S}}$, which is consistent with previous two-body decay measurements of $B^{0}\to K^{*\pm}K^{\mp}$~\cite{LHCb_kstk} and $B^{0}\to \bar{K}^{0}K^{*}(892)^{0}$~\cite{babar_kstk0,LHCb_kstks}.

\begin{figure}[htb]
\centering
\subfigure[$M_{K^{-}\pi^{+}}$]{
\includegraphics[width=0.3\textwidth]{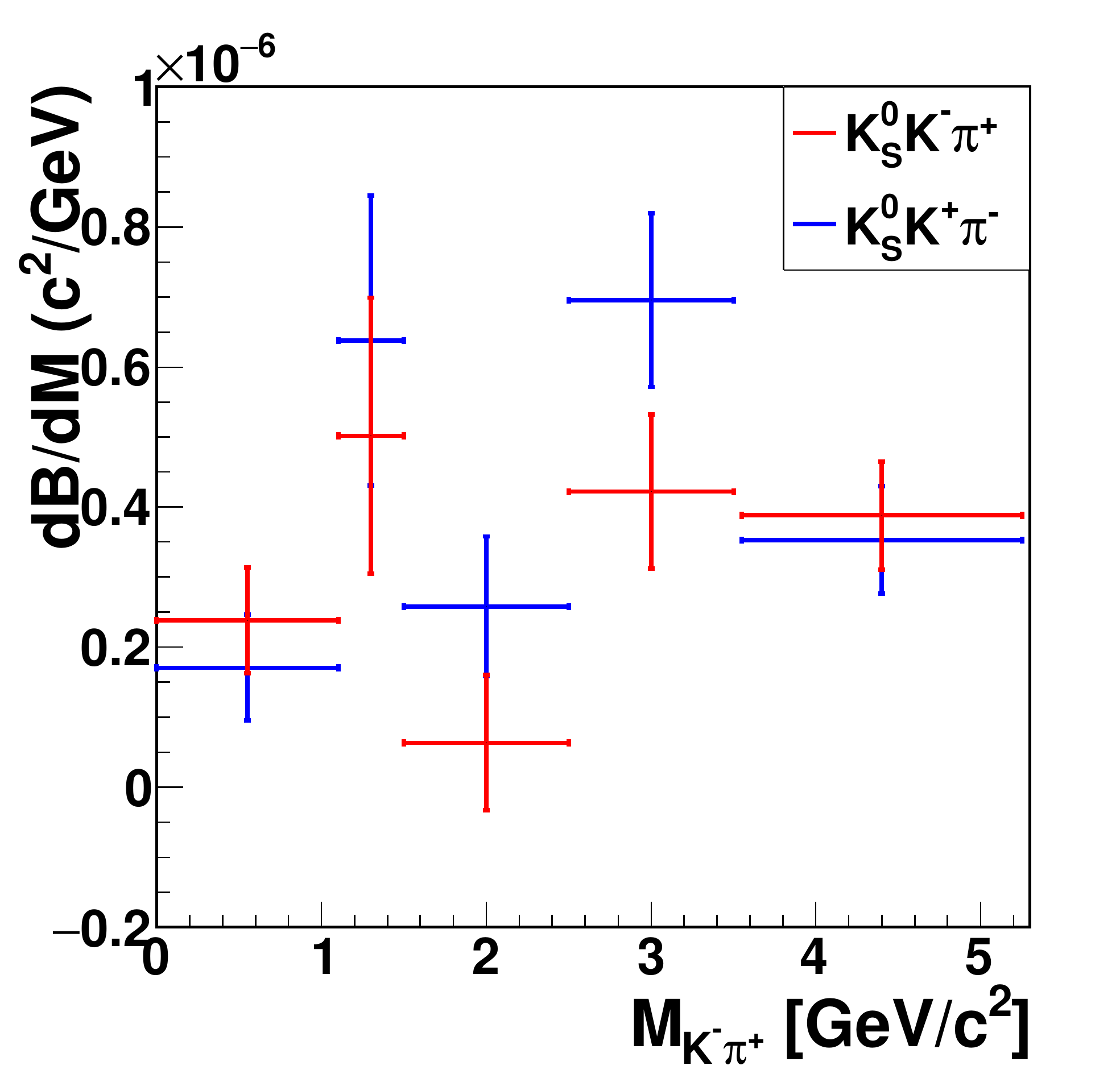}}
\subfigure[$M_{K^{-}K^{0}_{S}}$]{
\includegraphics[width=0.3\textwidth]{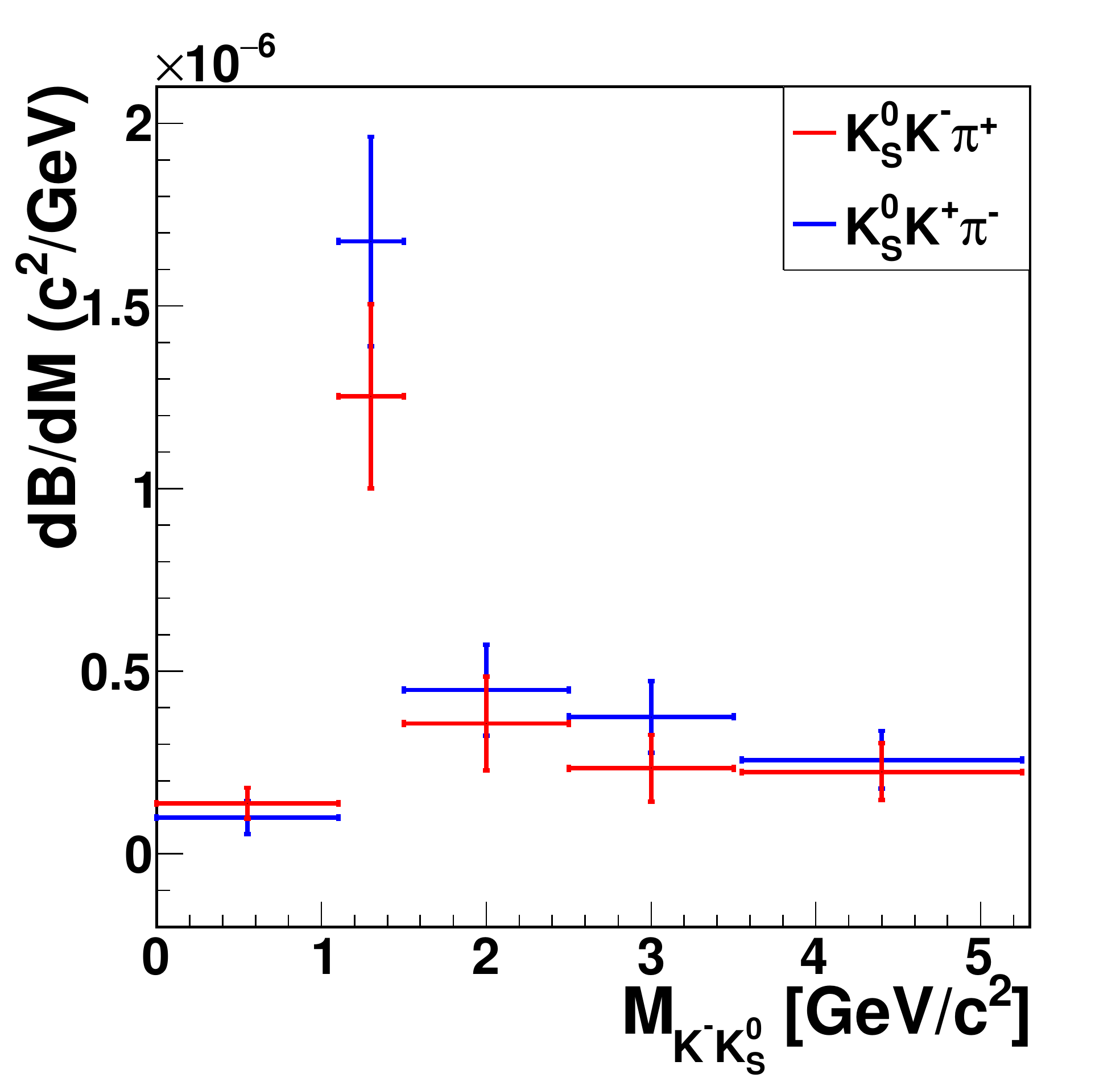}}
\subfigure[$M_{\pi^{+}K^{0}_{S}}$]{
\includegraphics[width=0.3\textwidth]{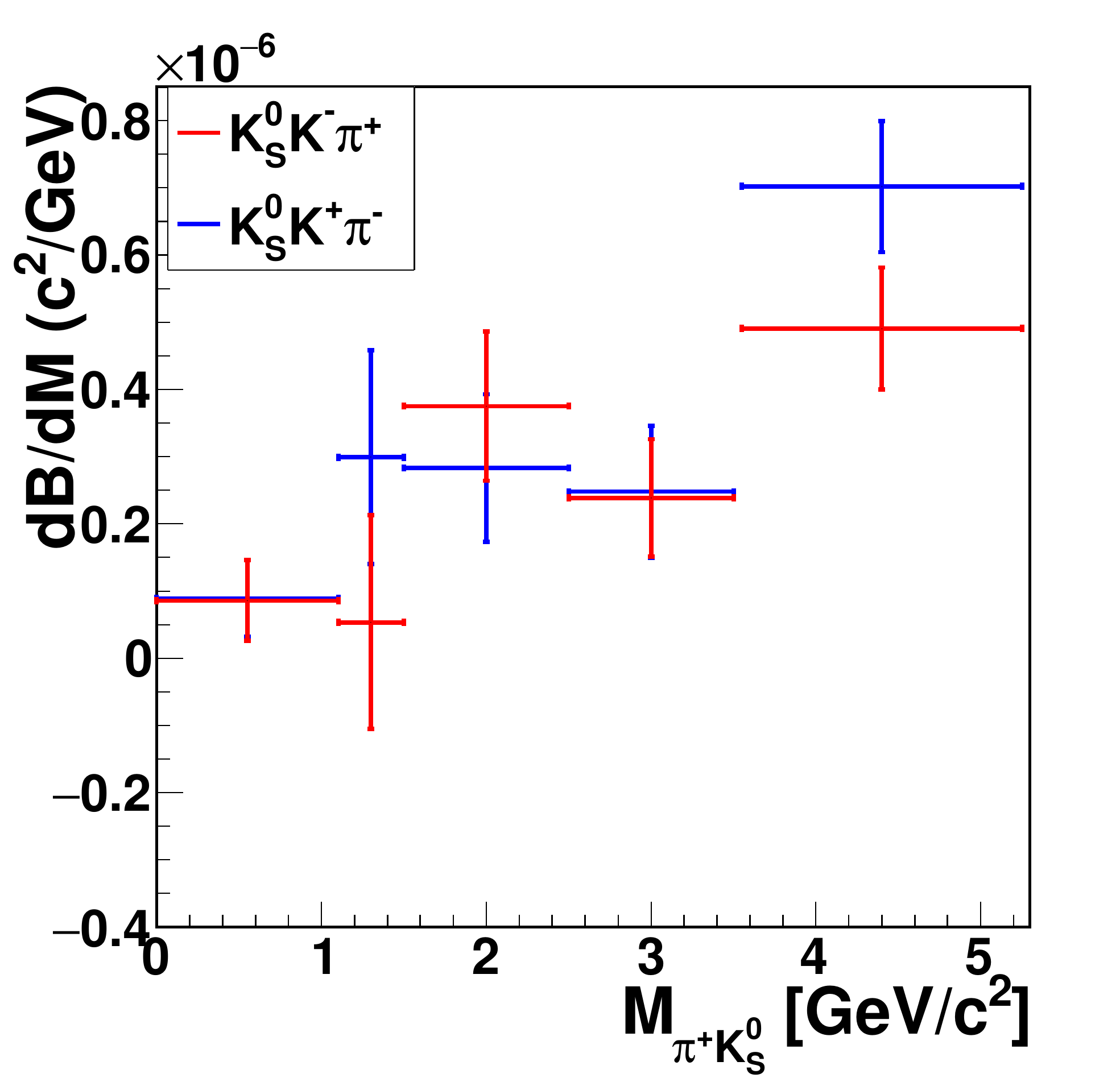}}
\caption{Differential branching fraction as functions of the $M_{K^{-}\pi^{+}}$, $M_{K^{-}K^{0}_{S}}$, and $M_{\pi^{+}K^{0}_{S}}$ for the two reconstructed $B$ final states: $K^{0}_{S}K^{-}\pi^{+}$ (points with red error bars) and $K^{0}_{S}K^{+}\pi^{-}$ (points with blue error bars).}
\label{fg:dbf_separate} 
\end{figure}

To investigate the localized final-state asymmetry, differential branching fractions separately for the $K^{0}_{S}K^{-}\pi^{+}$ and $K^{0}_{S}K^{+}\pi^{-}$ final states are shown in Fig.~\ref{fg:dbf_separate}.
Within each bin of the Dalitz variables, the results are consistent with no asymmetry. The details of differential branching fraction calculation in each bin are summarized in Table~\ref{tb:dbf}.

\begin{table}[t]
\begin{center}
\caption{Signal yields, efficiency, and differential branching fraction in each $M_{K^{-}\pi^{+}}$, $M_{K^{-}K^{0}_{S}}$, and $M_{\pi^{+}K^{0}_{S}}$ bin.}

\scriptsize
\begin{tabular}{c|ccc|cccc}
\hline \hline
 \multirow{2}{*}{($c^{2}$/GeV)} & \multirow{2}{*}{eff.} & \multirow{2}{*}{Yield} & \multirow{2}{*}{$d\mathcal{B}/dM~(10^{-7})$} & $K^{0}_{S}K^{-}\pi^{+}$ & $K^{0}_{S}K^{+}\pi^{-}$ & $K^{0}_{S}K^{-}\pi^{+}$ & $K^{0}_{S}K^{+}\pi^{-}$ \\
&  & & & yield & yield & $d\mathcal{B}/dM~(10^{-7})$ & $d\mathcal{B}/dM~(10^{-7})$ \\
\hline \hline
$M_{K^{-}\pi^{+}}$&&  &  & & &  &  \\
\hline
0$\sim$1.1 & 0.301 &  $69.2\pm18.0\pm3.0$ &$4.1\pm1.1\pm0.2$ & $40.3\pm12.7\pm1.7$ & $28.9\pm12.8\pm1.2$ & $2.4\pm0.7\pm0.1$ & $1.7\pm0.8\pm0.1$ \\
1.1$\sim$1.5 & 0.306 & $71.3\pm17.8\pm3.1$ & $11.4\pm2.8\pm0.5$ & $31.4\pm12.3\pm1.4$ & $39.9\pm12.9\pm1.7$ & $5.0\pm2.0\pm0.2$ & $6.4\pm2.1\pm0.3$ \\
1.5$\sim$2.5 & 0.289 & $47.5\pm20.5\pm2.0$ & $3.2\pm1.4\pm0.1$ & $9.4\pm14.3\pm0.4$ & $38.1\pm14.7\pm1.6$ & $0.6\pm1.0\pm0.0$ & $2.6\pm1.0\pm0.1$ \\
2.5$\sim$3.5 & 0.262 & $149.7\pm21.7\pm6.4$ & $11.2\pm1.6\pm0.5$ & $56.5\pm14.6\pm2.4$ & $93.2\pm16.1\pm4.0$ & $4.2\pm1.1\pm0.2$ & $7.0\pm1.2\pm0.3$ \\
$>$3.5 &0.237 &  $152.7\pm22.0\pm6.6$ & $7.4\pm1.1\pm0.3$ & $79.9\pm15.5\pm3.4$ & $72.8\pm15.5\pm3.1$ & $3.9\pm0.8\pm0.2$ & $3.5\pm0.8\pm0.2$ \\
\hline \hline
$M_{\pi^{+}K^{0}_{S}}$&&  &  & & &  &  \\
\hline
0$\sim$1.1 & 0.275 &  $27.1\pm 12.7 \pm1.2$ &$1.8\pm 0.8 \pm0.1$ & $13.3\pm 9.2 \pm0.6$ & $13.8 \pm8.7 \pm0.6$ & $0.9 \pm0.6\pm 0.0$ & $0.9\pm 0.6\pm 0.0$ \\
1.1$\sim$1.5 & 0.269 & $19.4\pm 12.4\pm 0.8$ & $3.5\pm 2.2 \pm0.2$ & $3.0\pm 8.8 \pm0.1$ & $16.5\pm 8.7 \pm0.7$ & $0.5\pm 1.6 \pm0.0$ & $3.0\pm 1.6\pm 0.1$ \\
1.5$\sim$2.5 & 0.252 & $84.8\pm 20.0\pm 3.6$ & $6.6\pm 1.5\pm 0.3$ & $48.3 \pm14.2 \pm2.1$ & $36.5 \pm14.1 \pm1.6$ & $3.8 \pm1.1\pm 0.2$ & $2.8 \pm1.1 \pm0.1$ \\
2.5$\sim$3.5 & 0.264 & $65.7 \pm17.6\pm 2.8$ & $4.9 \pm1.3\pm 0.2$ & $32.2 \pm11.7 \pm1.4$ & $33.4 \pm13.2 \pm1.4$ & $2.4 \pm0.9 \pm0.1$ & $2.5\pm 1.0 \pm0.1$ \\
$>$3.5 &0.283 &  $293.4 \pm31.5\pm 12.6$ & $11.9 \pm1.3\pm 0.5$ & $120.7 \pm21.7\pm 5.2$ & $172.7\pm 22.8\pm 7.4$ & $4.9\pm 0.9\pm 0.2$ & $7.0\pm0.9 \pm0.3$ \\
\hline \hline

$M_{K^{-}K^{0}_{S}}$&&  &  & & &  &  \\
\hline
0$\sim$1.1 & 0.245 &  $32.9\pm 8.5\pm 1.4$ &$2.4 \pm0.6\pm 0.1$ & $19.1 \pm5.8 \pm0.8$ & $13.7\pm 6.2\pm 0.6$ & $1.4\pm 0.4 \pm0.1$ & $1.0 \pm0.5 \pm0.0$ \\
1.1$\sim$1.5 & 0.258 & $154.6\pm 19.6 \pm6.6$ & $29.3 \pm3.7\pm 1.3$ & $66.1\pm 13.0\pm 2.8$ & $88.5 \pm14.7\pm 3.8$ & $12.5\pm 2.5 \pm0.5$ & $16.8 \pm2.8\pm 0.7$ \\
1.5$\sim$2.5 & 0.235 & $96.9 \pm21.3\pm 4.2$ & $8.1 \pm1.8 \pm0.3$ & $43.0\pm 15.3 \pm1.8$ & $53.9 \pm14.8 \pm2.3$ & $3.6 \pm1.3\pm 0.2$ & $4.5 \pm1.2 \pm0.2$ \\
2.5$\sim$3.5 & 0.267 & $83.4\pm 18.1 \pm3.6$ & $6.1\pm 1.3 \pm0.3$ & $32.1\pm 12.3 \pm1.4$ & $51.3 \pm13.2 \pm2.2$ & $2.4\pm 0.9\pm 0.1$ & $3.8\pm 1.0 \pm0.2$ \\
$>$3.5 &0.292 &  $122.6\pm 27.8\pm 5.3$ & $4.8 \pm1.1 \pm0.2$ & $57.2\pm 19.5 \pm2.5$ & $65.5\pm 19.9 \pm2.8$ & $2.3\pm 0.8 \pm0.1$ & $2.6\pm 0.8\pm 0.1$ \\
\hline \hline
\end{tabular}
\label{tb:dbf}
\end{center}
\end{table}

Sources of various systematic uncertainties in the branching fraction calculation are shown in Table~\ref{tb:sys}. The uncertainty due to the total number of $B\bar{B}$ pairs is 1.4\%. The uncertainty due to the charged-track reconstruction efficiency is estimated to be 0.35\% per track by using partially reconstructed $D^{*+}\to D^0 \pi^{+}$ with $D^0 \to \pi^+ \pi^- K^{0}_{S}$ events. The uncertainties due to $K^{\pm}$ and $\pi^{\pm}$ identification are obtained by the control sample study of $D^{*+}\to D^{0}\pi^{+}$ with $D^{0}\to K^{-}\pi^{+}$. The uncertainty due to the $K^{0}_{S}\to\pi^+ \pi^-$ branching fraction is based on the world average value $(69.2\pm0.05)\%$~\cite{PDG}. The uncertainty due to $K^{0}_{S}$ identification is estimated to be 1.6\% based on a $D^{*+}\to D^{0} \pi^{+}$, $D^{0}\to K^{0}_{S}\pi^{0}$ control sample~\cite{Ks_sys}. The uncertainty due to continuum suppression with the requirement on $C_{\mathrm{NN}}$ is obtained from a $B^{0}\to D^{-}\pi^{+}$ with a $D^{-}\to K^{0}_{S}\pi^{-}$ decay control sample. The uncertainty of the reconstruction efficiency is due to limited MC statistics. The uncertainty due to the fixed signal and background PDF shapes is estimated by the deviation of fitted signal yield when varying the parameters of the PDFs in different cases. For all the smoothed histograms, we vary the binning conditions of those histograms. For the other PDFs with fixed parameterization, the fixed parameters are randomized by using a Gaussian random number to repeat data fits with various parameter sets, and the uncertainty of the yield distribution is quoted. The uncertainty due to fit bias is obtained from an ensemble test on the fitter.

\begin{table}[htbp]
\begin{center}
\caption{Summary of systematic uncertainties on the branching fraction.}
\begin{tabular}{c|c}
\hline
\hline
Source & in \% \\ \hline
$N_{B\bar{B}}$ & 1.4 \\ 
Tracking & 0.7 \\ 
$K^{\pm}$ identification & 0.8 \\ 
$\pi^{\pm}$ identification & 0.8 \\ 
$\mathcal{B}(K^{0}_{S}\to\pi^{+}\pi^{-})$ & 0.1 \\ 
$K^{0}_{S}\to\pi^{+}\pi^{-}$ identification & 1.6 \\ 
Continuum suppression with NN & 2.1 \\ 
Reconstruction efficiency (MC statistics) & 0.1 \\ 
Signal PDF & 2.7 \\ 
Background PDF & 0.4 \\ 
Fit bias & 0.4 \\ \hline
Total & 4.3 \\
\hline
\hline 
\end{tabular}
\label{tb:sys}
\end{center}
\end{table}

Sources of various systematic uncertainties on $\mathcal{A}$ are listed in Table~\ref{tb:sys_acp}. The uncertainty due to $K^{\pm}$ and $\pi^{\pm}$ detection bias are obtained by control sample studies of $D^{+}\to \phi\pi^{+}$ and $D^{+}_{s}\to\phi \pi^{+}$~\cite{phipi}, and $D^{+}\to K^{0}_{S}\pi^{+}$~\cite{kspi}, respectively.  The uncertainties due to the fixed signal and background PDF shapes are treated in the same way as those in the uncertainty on the branching fraction. The systematic uncertainties due to PDF's are also estimated from the deviation of the fitted value of $\mathcal{A}$ with varying the conditions of those PDFs in different cases.

\begin{table}[htbp]
\begin{center}
\caption{Summary of systematic uncertainties on $\mathcal{A}$.}
\begin{tabular}{c|c}
\hline
\hline
Source & in \% \\ \hline
Detector bias & 0.6 \\ 
Signal PDF & 2.7 \\ 
Background PDF & 0.9 \\ \hline
Total & 2.9 \\
\hline
\hline 
\end{tabular}
\label{tb:sys_acp}
\end{center}
\end{table}

In conclusion, we have performed a measurement of the branching fraction and asymmetry $\mathcal{A}$ of the $\bar{B}^{0}\to K^{0}_{S}K^{\mp}\pi^{\pm}$ decay based on a data sample of 711 fb$^{-1}$ collected by Belle. We obtain a branching fraction of $(3.60\pm0.33\pm0.15)\times10^{-6}$ and an $\mathcal{A}$ of $(-8.5\pm8.9\pm0.2)\%$, where their first uncertainty is statistical and the second is systematic. The measured $\mathcal{A}$ value is consistent with no asymmetry. Hints of peaking structures are seen in the regions $M^{2}_{K^{-}K^{0}_{S}}<$ 2 GeV$^{2}$/$c^{4}$ and 7 GeV$^{2}$/$c^{4}$ $<M^{2}_{\pi^{+}K^{0}_{S}} <$ 23 GeV$^{2}$/$c^{4}$ in the Dalitz plot. 
A cross-check was performed by calculating the differential branching fraction after projecting onto each Dalitz variable, and hints of peaking structures are found near 1.2 GeV/$c^{2}$ in $M_{K^{-}K^{0}_{S}}$ and around 4.2 GeV/$c^{2}$ in $M_{\pi^{+}K^{0}_{S}}$ when compared to the phase space MC. No obvious $K^{*}$ structure is seen either in low $M_{K^{-}\pi^{+}}$ and $M_{\pi^{+}K^{0}_{S}}$ spectra, which are also consistent with the BABAR and LHCb results~\cite{babar_kstk0,LHCb_kstk,LHCb_kstks}. No localized final-state asymmetry is observed. In the near future, experiments with large data sets such as Belle II and LHCb can provide a more detailed analysis exploiting the full Dalitz plot to search for intermediate resonances and localized final-state asymmetry.  

We thank the KEKB group for the excellent operation of the
accelerator; the KEK cryogenics group for the efficient
operation of the solenoid; and the KEK computer group,
the National Institute of Informatics, and the 
PNNL/EMSL computing group for valuable computing
and SINET4 network support.  We acknowledge support from
the Ministry of Education, Culture, Sports, Science, and
Technology (MEXT) of Japan, the Japan Society for the 
Promotion of Science (JSPS), and the Tau-Lepton Physics 
Research Center of Nagoya University; 
the Australian Research Council;
Austrian Science Fund under Grant No.~P 22742-N16 and P 26794-N20;
the National Natural Science Foundation of China under Contracts 
No.~10575109, No.~10775142, No.~10875115, No.~11175187, No.~11475187
and No.~11575017;
the Chinese Academy of Science Center for Excellence in Particle Physics; 
the Ministry of Education, Youth and Sports of the Czech
Republic under Contract No.~LG14034;
the Carl Zeiss Foundation, the Deutsche Forschungsgemeinschaft, the
Excellence Cluster Universe, and the VolkswagenStiftung;
the Department of Science and Technology of India; 
the Istituto Nazionale di Fisica Nucleare of Italy; 
the WCU program of the Ministry of Education, National Research Foundation (NRF) 
of Korea Grants No.~2011-0029457,  No.~2012-0008143,  
No.~2012R1A1A2008330, No.~2013R1A1A3007772, No.~2014R1A2A2A01005286, 
No.~2014R1A2A2A01002734, No.~2015R1A2A2A01003280 , No. 2015H1A2A1033649;
the Basic Research Lab program under NRF Grant No.~KRF-2011-0020333,
Center for Korean J-PARC Users, No.~NRF-2013K1A3A7A06056592; 
the Brain Korea 21-Plus program and Radiation Science Research Institute;
the Polish Ministry of Science and Higher Education and 
the National Science Center;
the Ministry of Science and Higher Education of Russian Federation, Agreement 14.W03.31.0026;
the Slovenian Research Agency;
Ikerbasque, Basque Foundation for Science and
the Euskal Herriko Unibertsitatea (UPV/EHU) under program UFI 11/55 (Spain);
the Swiss National Science Foundation; 
the Ministry of Education and the Ministry of Science and Technology of Taiwan;
and the U.S.\ Department of Energy and the National Science Foundation.
This work is supported by a Grant-in-Aid from MEXT for 
Science Research in a Priority Area (``New Development of 
Flavor Physics'') and from JSPS for Creative Scientific 
Research (``Evolution of Tau-lepton Physics'').


\begin{thebibliography}{99}
\bibitem{charge_conjugate} Throughout this paper, inclusion of charge-conjugate decay modes is implied unless otherwise stated.
\end{thebibliography}
\end{document}